\documentclass[twocolumn,amsmath,amssymb]{revtex4}
\usepackage{graphicx,subfigure}
\usepackage{array}

\newcommand{\ave}[1]{\langle #1 \rangle}
\newcommand{\bra}[2][]{\mathinner{\langle #2|}_{#1}}
\newcommand{\ket}[2][]{\mathinner{|#2\rangle}_{\hspace{-0.1em}#1}}
\newcommand{\defas}{:=}
\DeclareMathOperator{\tr}{Tr}
\newcommand{\abs}[1]{|#1|}
\newcommand{\bigO}[1]{\mathcal{O}\left(#1\right)}

\newcommand{\PreserveBackSlash}[1]{\let\temp=\\#1\let\\=\temp}
\let\PBS=\PreserveBackSlash
\newcommand{\RR}{\PBS\centering\hspace{0pt}}

\begin{document}

\title{Minimum Energy-Surface Required by Quantum Memory Devices}

\author{Wim van Dam}
\email[Email: ]{vandam@cs.ucsb.edu}
\affiliation{%
Department of Computer Science, Department of Physics, University of California, Santa Barbara, CA 93106-5110, USA
}

\author{Hieu D.\ Nguyen}
\email[The authors are listed in alphabetical order; HDN is the primary and corresponding author. Email: ]{hdn@physics.ucsb.edu}
\affiliation{%
Department of Physics, University of California, Santa Barbara, CA 93106-5110, USA
}%

\date{\today}\begin{abstract}
We address the question what physical resources are required and  
sufficient to store classical information. While there is no lower  
bound on the required energy or space to store information, we find  
that there is a nonzero lower bound for the product $P =  
\ave{E} \ave{r^2}$ of these two resources. Specifically, we prove that any physical system of mass $m$ and  
$d$ degrees of freedom that stores $S$ bits of information  
will have lower bound on the product $P$ that is proportional  
to $d^2/m(\exp(S/d)-1)^2$. This result is obtained in a non-relativistic, quantum mechanical setting and it is independent from earlier thermodynamical results such as the Bekenstein bound on the entropy of black holes.
\end{abstract}

\pacs{pacs numbers}
\maketitle

\section{Introduction.} 
Although information may seem abstract and elusive, it has made a number of surprisingly concrete connections to physics. Originally prompted by Maxwell's demon, the link of information to thermodynamics has been made lucid by the works of Bennett and Landauer (see \cite{dem90} for complete coverage). Physicists and computer scientists alike have come to recognize that information does not exist as an independent entity but is encoded in physical devices, and, therefore,  ``Information is physical'' has become a new mantra \cite{lan96}. Moreover, the technological trend from the last few decades continually has been to compress more information in less space \cite{hdd}. Due to Moore's law we sense that we are near the end of the classical regime and entering into the quantum regime where exciting possibilities and challenges abound. Thus, to understand the ultimate limits to information storage based on the quantum laws of physics may be of intellectual as well as practical significance. In this paper we seek a general and robust result concerning information storage using non-relativistic quantum theory. Abstracting away the details of hardware, we compute the absolute minimal amount of physical resources required to store information, and the result further illustrates its physical nature.

Typically, information is stored by preparing a device in a number of different stationary states corresponding to different messages. The device is effectively described by a mixed state of a certain entropy, and looking at a number of simple devices such as particle in a box, harmonic oscillator, and the hydrogen atom, we notice an interesting interplay among energy, surface area, mass, and entropy (see Table~\ref{table}). Keeping the constituent particles' masses the same, one could save on energy or size but never both. For instance, the particle-in-a-box model requires the same amount of space regardless of the number of states used in the encoding but the energy increases with the number of states. The hydrogen atom is an example of the opposite extreme. There is an infinite number of states in a finite energy window, roughly equal to $13.6$ eV, but the spatial size of the orbitals increases without bound. However, if we could change the mass, then both energy and space can be reduced arbitrarily. Since a number of variables could affect the cost, it is necessary to hone in on a precise notion of the cost of information storage. We will do so in the next section and determine the optimal device based on a number of clearly stated plausible hypotheses. Then, we proceed to compute the minimum cost and prove a lemma which is needed in the calculation. Our main result is a Heisenberg-like uncertainty principle governing energy, space, and entropy. We conclude with a comparison to rough estimates existing in the literature as well as some thoughts upon its implications for future technological advances.
 \begin{table}\label{table}
  \begin{center}
\begin{tabular}%
{>{\RR}m{3cm}%
c%
>{\RR}m{1.3cm}%
>{\RR}m{1.3cm}%
>{\RR}m{1.5cm}}
System & & Energy & Surface Area & Product\\\hline~\\
Particle of mass $m$ in a 1d box of width $L$& &  $\frac{\hbar^2 \pi^2}{m L^2}\ave{n^2}$ & $L^2$ & $\frac{\hbar^2}{m}\ave{n^2}$ \\ ~\\
Harmonic oscillator of mass $m$ and natural frequency $\omega$ & & $\hbar \omega \ave{n}$ & $\frac{\hbar}{m \omega}\ave{n}$ & $\frac{\hbar^2}{m}\ave{n}^2$ \\  ~\\
Electronic states of Hydrogen atom & & $\frac{e^4 m}{\hbar^2}\ave{\frac{1}{n^2}}$ & $\frac{\hbar^4}{m^2 e^4}\ave{n^4}$ & $\frac{\hbar^2}{m}\ave{n^4}\ave{1/n^2}$ \\~\\
   \end{tabular}
 \caption{A system is prepared in different stationary states, labelled by $n$, according to a probability distribution whose Shannon entropy is the desired amount of encoded information. Quoted are order-of-magnitude estimates of the average energy and surface area, denoted by $\ave{\cdot}$, for a number of simple systems \cite{sakurai}. Averages such as $\ave{n}$, $\ave{n^2}$, and $\ave{n^4}$ grow with increasing entropy. Therefore, although energy and surface area show different behaviours for different systems, their products invariably grow with entropy. }
  \end{center}
 \end{table}
\section{Definition of cost}
Consider storing information in a device described by the Hamiltonian
\begin{equation}
H_{\text{dev}} = -\frac{\hbar^2}{2m} \sum_{i=1}^{d}\frac{d^2}{dx_{i}^{2}} + V(\vec{x}),
\end{equation}
by preparing it in a stationary (energy eigenstate) $\Psi_{\alpha}$ with probability $p_{\alpha}$. The amount of information thus encoded is
\begin{equation}
\label{eq:entropy}
S = -\sum\limits_{\alpha}p_{\alpha}\log p_{\alpha}
\end{equation}
and the device will be in the state
\begin{equation}
\rho_{\text{dev}} = \sum\limits_{\alpha}p_{\alpha}\ket{\Psi_{\alpha}}\bra{\Psi_{\alpha}}.
\end{equation}
We consider the average energy and size associated with such a mixed state the cost in physical resources required by the device. More specifically, let $\tr{\rho_{\text{dev}}r^2}$
be the measure of spatial cost where 
\begin{equation}
r^2 \defas \sum\limits_{i=1}^{d} x_{i}^{2}.
\end{equation}
To prevent arbitrary shifts due to displacements of the coordinate origin, we require
for all $i\in\{1,\dots,d\}$ that 
$\tr{\rho_{\text{dev}} x_i} = 0$. 
Moreover, let $\tr{\rho_{\text{dev}}H_{\text{dev}}}$ be a measure of the cost in energy, and similarly, to prevent meaningless shifts in the Hamiltonian, we require the ground state (g.s.) energy to be zero,
\begin{equation}
\label{eq:pot-const}
\langle H_{\text{dev}} \rangle_{\text{g.s.}} = 0
\end{equation}
The cost of information storage specific to using $H_{\text{dev}}$ is the joint product of energy and space
\begin{equation}
\label{eq:dev-cost}
C_{\text{dev}} \defas \min_{\rho_{\text{dev}}}\tr{\rho_{\text{dev}}H_{\text{dev}}} \tr{\rho_{\text{dev}}r^2},
\end{equation}
subject to the constraint of Equation (\ref{eq:entropy}). Since the cost in Equation (\ref{eq:dev-cost}) is specific to a device, to find the cost of information storage we need to minimize over all devices having the same number of degrees of freedom and particles' mass as follows:
\begin{equation}
\label{eq:info-cost}
P \defas \min_{V(x)} C_{\text{dev}}
\end{equation}
over potentials that satisfy (\ref{eq:pot-const}). 

We choose to express the intuitive notion of cost in energy and space by considering linear power in energy and quadratic in length, \eqref{eq:dev-cost}. One could potentially capture the same intuition through other powers of the average $H$ and $r$ such as $
\ave{H^{s_1}} \ave{r^{s_2}}$. Putting the exponents inside or outside of the average may change the end result slightly but should convey the same physics. Since the only dimensionful parameters in our problem are $\hbar$ and $m$, if the cost is measured in units of $J^{s_1} l^{s_2}$, where $J$ and $l$ stand for Joules and meters, then the answer must be expressed in powers $t_1$ of $\hbar$ and $t_2$ of $m$ such that
\begin{equation}
(J \cdot s)^{t_1} kg^{t_2} = J^{s_1} l^{s_2},
\end{equation} 
where $s$ and $kg$ stand for seconds and kilograms. Solving this equation leads to these constraints on the exponents: $t_1 = 2 s_1$, $t_2 = -s_1$, and $s_2 = 2 s_1$.  
The first two equations guarantee that other powers of energy and space admit a solution and therefore are also valid, but, as implied by the last equation, they all are powers of the basic combination of linear in energy and quadratic in space.
\section{Optimal potential}
For the purpose of computing a lower bound of $P$, we can relax the condition that the mixed state be diagonal in the eigenstates of $H_{\text{dev}}$ and consider instead the following functional of $V(\vec{x})$
\begin{equation}
\label{eq:defn-c}
C[V(\vec{x})] \defas \min\limits_{\rho} \left\{ \tr{\rho H_{\text{dev}}} \tr{\rho r^2}\right\},
\end{equation}
 minimized over all density matrices satisfying the entropy condition
\begin{equation}
\label{eq:entr-const}
S = -\tr{\rho \log \rho}.
\end{equation}
Obviously, for a given $V(x)$,
\begin{equation}
C_{\text{dev}} \geq C[V(x)]
\end{equation}
Hence,
\begin{equation}
\label{ineq}
P \geq  \min\limits_{V(x)} C[V(x)] 
\end{equation}
with $V(x)$ subject to the same constraint of Equation $\eqref{eq:pot-const}$. Let us start by observing that $C[V(\vec{x})]$ is rotationally invariant. Under a coordinate rotation $\vec{y} = R \vec{x}$, with $R \in SO(d)$, we have 
\begin{align}
& C[V(R \vec{x})] \\
&= \min\limits_{\rho} \left\{ \tr{\rho \left(-\frac{\hbar^2}{2m} \sum_{i=1}^{d}\frac{d^2}{dx_{i}^{2}} + V(R \vec{x})\right)} \tr{\rho r^2}\right\}\\
 &=  \min\limits_{\rho} \left\{ \tr{\rho \left(-\frac{\hbar^2}{2m} \sum_{i=1}^{d}\frac{d^2}{dy_{i}^{2}} + V(\vec{y})\right)} \tr{\rho r^2}\right\}\\
 &= C[V(\vec{x})]
\end{align} 
The first equality comes from applying the definition in Equation $~\eqref{eq:defn-c}$; the second comes from using the $y$-coordinates instead of $x$ and rotational invariance of kinetic energy and $r^2$; finally, the third comes from applying the definition again. Thus, if we assume a minimizing potential $V(\vec{x})$ exists, then $C[V(R \vec{x})]$ must also be minimal by rotational invariance. Moreover, if we assume the optimal potential to be also unique, then $V(R \vec{x}) = V(\vec{x})$. Hence, by assuming existence and uniqueness, we conclude the optimal potential must be a function of only the hyper-radius $r$. 
As we try to find the potential which minimizes $C[V(\vec{x})]$, we have in the problem only the parameters  $\hbar$, $m$, and $r$ to construct our optimal potential. By dimensional analysis one is led to conjecture the solution to be $V(r) = -W\frac{\hbar^2}{2m r^2}$, where $W$ is a dimensionless number possibly depending on $S$. The larger $W$, that is the more attractive the potential, the smaller the orbitals and total energy, thus allowing us to save on the total cost without compromise. However, the attractive inverse square potential for $W > (1-d/2)^2$ is unstable as it has negative infinity ground state energy, .i.e. ``fall to the center'' \cite{kw74, gr93, sh31, landau}. Thus, the optimal potential occurs at the maximum allowed value of $W$ before it collapses, namely 
\begin{equation}
H_{\text{opt}} = -\frac{\hbar^2}{2m} \sum_{i=1}^{d}\frac{d^2}{dx_{i}^{2}} - \Big(1-\frac{d}{2}\Big)^2\frac{\hbar^2}{2m r^2}
\end{equation}
It remains a conjecture for the rest of our paper that $H_{\text{opt}}$ yields the smallest cost possible.
\section{Minimum cost}
Since $H_{\text{opt}}$ minimizes $C[V(x)]$, by Equations (\ref{eq:defn-c}) and (\ref{ineq}), we have
\begin{equation}
\label{ineq2}
P \geq \min\limits_{\rho} \left\{ \tr{\rho H_{\text{opt}}} \tr{\rho r^2}\right\}.
\end{equation}
To evaluate the right hand side, let us introduce $\kappa > 0$ and define
\begin{equation}\label{eq:costsum}
\tilde{C} \defas  \min\limits_{\rho}\tr{\rho\left(H_{\text{opt}} + \frac{\kappa}{2}r^2\right)} 
\end{equation}
with $\rho$ subject to the entropy constraint Equation $~\eqref{eq:entr-const}$. The auxiliary variable $\kappa$ allows us to convert the cost in space into an energy quantity so that in a single expression, $\tilde{C}$ is the joint costs of energy and space in the form of a sum. As to be shown in the subsection, we have
\begin{equation}
\label{lemma}
\tilde{C} \geq \hbar \sqrt{\frac{\kappa}{m}} d (\exp{(S/d)}-1)
\end{equation}
Hence, for the mixed state obtaining the minimum of the right hand side in Equation (\ref{eq:costsum}), we have
\begin{equation}
\tr{\rho H_{\text{opt}}} \geq \hbar \sqrt{\frac{\kappa}{m}} d(\exp{(S/d)}-1)-\frac{\kappa}{2} \tr{\rho r^2}, 
\end{equation}
which yields
\begin{align*}
&\tr{\rho H_{\text{opt}}}\tr{\rho r^2} \geq \\
&\sqrt{\kappa}\left(\hbar \frac{d}{\sqrt{m}}(\exp{(S/d)}-1)-\frac{\sqrt{\kappa}}{2}\tr{\rho r^2}\right)\tr{\rho r^2}
\end{align*}
for all non-negative values of $\kappa$. Since the right hand side is quadratic in $\sqrt{\kappa}$, at the critical $\kappa$ the inequality yields
\begin{equation}
\tr{\rho H_{\text{opt}}}\tr{\rho r^2}\geq \frac{\hbar^2}{2m}d^2 (\exp{(S/d)}-1)^2,
\end{equation}
which is now free of the introduced variable $\kappa$. Because of inequality (\ref{ineq2}), the cost of storing information satisfies the bound
\begin{equation}\label{eq:mainresult}
P \geq \frac{\hbar^2}{2m}d^2 (\exp{(S/d)}-1)^2,
\end{equation}
which is our main result.
\subsection{Central Lemma} 
We now proceed to prove lemma (\ref{lemma}). Simplifying the dimensionful quantities in $\tilde{C}$ yields
\begin{equation}
\label{eq:dim-less}
\tilde{C} =  \hbar\sqrt{\kappa/m} \min\limits_{\rho}\tr{\rho H},
\end{equation}
where
\begin{equation}
H  \defas \frac{1}{2} \sum_{i=1}^{d}-\frac{d^2}{dq_{i}^{2}} -\frac{W}{q^2} + q^2
\end{equation}
is dimensionless as is $\vec{q} \defas \vec{x}(m \kappa)^{1/4}$. Let 
\begin{equation}
\rho = \sum\limits_{i}p_{i}\ket{\Psi_{i}}\bra{\Psi_{i}}
\end{equation}
with eigenvalues $p_i$ and $\Psi_{i}$ orthogonal and not necessarily eigenstates of $H$. Thus, we would like to minimize 
$$
\sum\limits_{i}p_{i} \left|\bra{\Psi_{i}}H\ket{\Psi_{i}}\right|^2
$$
by varying $\ket{\Psi_{i}}$ and $p_i$ subject to the entropy condition (\ref{eq:entr-const}). Obviously, we would have the smallest sum possible if corresponding to the highest $p_i$ we have the smallest $\left|\bra{\Psi_{i}}H\ket{\Psi_{i}}\right|^2$ possible. Namely, it should be the ground state of $H$. Similarly, the next most frequently used $\Psi_{i}$ should be that which yields the next least expectation value of $H$. That is, it should be the first excited state. Continuing this argument, we see that $\Psi_{i}$ are eigenstates of $H$ with some yet to be determined distribution $p_i$. Optimization over $p_i$ with the constraints of entropy and normalization can be treated with Lagrange multipliers in a manner identical to that typically found in textbooks, thus resulting in the Boltzmann distribution.
In other words, the minimizing density matrix of Equation (\ref{eq:dim-less}) is diagonal in the eigenstates of $H$ with $p_i \propto \exp{(-\beta E_i)}$ with $E_i$ being eigen-energies of $H$ and $\beta$ a constant depending on $S$. The spectrum of $H$ can be computed by casting the Laplacian into hyper-spherical coordinates and solving the ensuing hypergeometric differential equation as in \cite{kw74}. Eigenstates are characterized by two quantum numbers $n, l = 0, 1, 2, \ldots$ and have energy
\begin{equation}
E(n,l) = 2n+ \sqrt{l(l+d-2)},
\end{equation}
with degeneracy
\begin{equation}
g(l) = \frac{(d+2l-2)(d+l-3)!}{l!(d-2)!}
\end{equation}
Thus, Equation (\ref{eq:dim-less}) becomes
\begin{equation}
\tilde{C} = \hbar\sqrt{\kappa/m} \frac{1}{Z}\sum\limits_{n=0}\sum\limits_{l=0} \exp{(-\beta E(n,l))} g(l) E(n,l)
\end{equation} 
where
\begin{equation}
\label{eq:part-fxn}
Z \defas \sum\limits_{n = 0}\sum\limits_{l = 0}g(l) \exp{(-\beta E(n,l))}.
\end{equation}
Moreover, the separation of $n$ and $l$ in the eigen-energy equation allows the partition function to be factorized
\begin{equation}
\label{eq:part_sep}
Z = \underbrace{\sum\limits_{n = 0} \exp{(-2\beta n)}}_{Z_n}\underbrace{\sum\limits_{l = 0}\exp{(-\beta\sqrt{l(l+d-2)})}g(l)}_{Z_l}
\end{equation}
Thus, although in Equation (\ref{eq:dim-less}) $\tilde{C}$ in our context means the cost of energy and space, it can also be interpreted as the internal energy of a thermodynamic system composed of two uncoupled sub-systems $n$ and $l$. Let $U_n$, $U_l$, $S_n$, and $S_l$ be internal energies and entropies of the $n$ and $l$ subsystems respectively, and we have
\begin{align}
\tilde{C} &= \hbar \sqrt{\frac{\kappa}{m}}\left(U_n+ U_l \right) \quad \mbox{and}\\
S &= S_n+ S_l, 
\end{align}
Because $Z_n$, being a geometric sum, can be computed in closed form, we get the following exact results for the thermodynamic quantities
\begin{align}
\beta &= \frac{1}{2}\log \left(1+\frac{2}{U_n}\right)\label{eq:betaUn}\\
S_n &=  \log \left(1+\frac{U_n}{2}\right) + \frac{U_n}{2}\log \left(1+\frac{2}{U_n}\right).
\end{align}
While the $n$ sum is trivial, we can make progress with the $l$-sum only when making approximations in certain regimes. In the following we will work in the limits $d \gg 1$ and $\beta \ll 1$, which corresponds to high entropy. Using Stirling's formula and the method of steepest descent to compute the integral version of the discrete sum, we get 
\begin{equation}\label{eq:part_approx}
Z_l \approx 2\beta^{-d+1}
\end{equation}
with the error governed by 
\begin{equation}
\abs{\log Z_l - \log 2\beta^{-d+1}} = \bigO{\frac{\beta}{d}} \quad \text{(See figure 1)}
\end{equation}
By the canonical equations yielding internal energy and entropy from the partition function,  we then obtain
\begin{align}
U_l &= \frac{d-1}{\beta} + \bigO{\frac{1}{d}}  \label{eq:Ul} \\
S_l &= (d-1) \left(\log \frac{U_l}{d-1}+1 \right) + \bigO{\frac{\beta}{d}} \label{eq:Sl}
\end{align}
Since we work in the regime $\beta \ll 1$, Equation $~\eqref{eq:betaUn}$ implies $U_n \gg 1$. Hence, keeping only the dominant terms in the $n$-subsystem, we obtain
\begin{align}
U_n & = \frac{1}{\beta} -1 +\bigO{\beta}  \label{eq:Un}\\
S_n & = \log \left(1+ \frac{U_n}{2}\right) + \bigO{1} \label{eq:Sn}
\end{align}
When we combine the above results for the separate sub-systems and keep only the dominant terms, we get
\begin{align}
\tilde{C} &= \hbar \sqrt{\frac{\kappa}{m}}\left(d U_n - (d-1)+\bigO{1/d}\right) \quad \mbox{and}\\
S &= d \log U_n + \bigO{1}, 
\end{align}
We now obtain the equation relating the amount of encoded information to the associated cost,
\begin{equation}
\frac{\tilde{C}}{d} = \hbar \sqrt{\frac{\kappa}{m}} \left(\exp{(S/d)} - \frac{d-1}{d}+ \bigO{1/d^2}\right),
\end{equation}
which yields lemma (\ref{lemma}) in the limit of large $d$.
\begin{figure}
\subfigure[Error]{%
\includegraphics[scale=0.5]{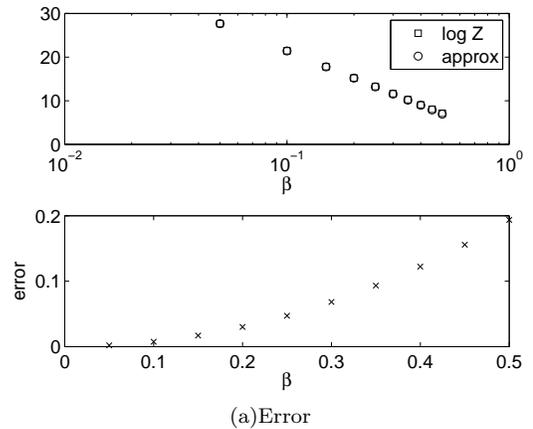}
}%
\caption{We computed $\log Z$ in two different ways for $d=10$. The $\log Z$-plot is the result of carrying out the sum in Equation $~\eqref{eq:part_sep}$ to $1600$ terms at which it seems stable. The approximation-plot is Equation $~\eqref{eq:part_approx}$ resulting from the method of steepest descent. The error plot shows the difference $\log Z - \log 2\beta^{-d+1}$, which is of the order $\beta/d$.}
\end{figure}

\section{Conclusion.}
Using non-relativistic quantum theory, we find the minimal physical resources needed to store information based on a number of plausible hypotheses. Defining the cost in the form of a product of energy and space, we proceed to find the interaction potential that minimizes the cost when storing a given amount of information. We assume such a potential exists and is rotationally invariant, which also follows if assuming uniqueness of the optimal potential. As our main result we arrive at Equation $~\eqref{eq:mainresult}$, which provides a lower bound to any device's energy and surface area when it encodes $S$ amount of information in $d$ degrees of freedom in the limits $d, S \gg 1$. According to our intuition we expect information to require no energy and space, and this only appears so because the factor $\hbar/m$ is extremely small in the classical regime. For comparison with some existing rough estimates in the literature, let us consider a spherical memory device made up of ordinary matter of total mass $1$ kilogram and volume $1$ litter. Since the energy in our result is not the rest-mass energy but the excitation energies relative to the ground, let us take the energy to be the ionization energy of typical atoms, which is $10$ eV. Thus, our memory looks like a soup of many electrons at a macroscopic distance from a positively charged center. Our inequality yields $S/d \approx 20$, about $20$ bits per atom, which agrees with Bekenstein's estimate for electronic levels but falls far short compared to $10^6$ when energy is taken to be the rest-mass energy \cite{bek84}. Hence, the whole device can hold about $10^{31}$ bits, which is the same as Lloyd's ultimate laptop, \cite{llo00, llo02}. Moreover, the cost is proportional to the number of degrees of freedom but exponential in the information density, $S/d$, thus showing analog storage to be much more expensive than digital. Given a device storing one bit of information, analog storage would keep the same device and use its remaining internal states to store additional bits, and this is shown to be exponentially more expensive than bringing in more copies of the same device, which is digital storage. Our current state-of-the-art technology stores gigabytes of information in a macroscopic device consisting of an Avogadro number of entities; that is, we are operating at the low density regime, $S/d \approx 10^{-14}$. As we try to create smaller devices with greater information capacity, we are pushing $S/d$ to greater values. However, our result implies this endeavour will become infeasible due to the exponential scaling.
\paragraph*{Acknowledgments.}
This material is based upon work supported by the  
National Science Foundation under Grant No. 0917244.

\end{document}